\documentclass[runningheads]{llncs}
\usepackage{graphicx}
\usepackage{subcaption}
\usepackage{color}
\usepackage{xcolor}
\usepackage{amsfonts}
\usepackage{multirow}
\usepackage{booktabs}
\usepackage{enumitem}
\usepackage{amsmath}
\usepackage{balance}
\usepackage{subcaption}
\usepackage{ulem}
\usepackage{hyperref} 
\usepackage{comment}
\usepackage{orcidlink}

\begin{document}

\title{Interpret3C: Interpretable Student Clustering Through Individualized Feature Selection}

\titlerunning{Interpret3C}

\author{Isadora Salles\orcidlink{0000-0001-8913-6294} \and
Paola Mejia-Domenzain*\orcidlink{0000-0003-1242-3134} \and
Vinitra Swamy*\orcidlink{0000-0002-6840-5923}\and \\
Julian Blackwell\orcidlink{0009-0004-5450-2759} \and
Tanja Käser\orcidlink{0000-0003-0672-0415}
}

\authorrunning{I. Salles et al.}
\institute{EPFL, Switzerland\\
\email{isadorasalles@dcc.ufmg.br, paola.mejia@epfl.ch, vinitra.swamy@epfl.ch, julian.blackwell@epfl.ch, tanja.kaeser@epfl.ch}}

\def\thefootnote{*}\footnotetext{Equal contribution.}

\maketitle             
\begin{abstract}

Clustering in education, particularly in large-scale online environments like MOOCs, is essential for understanding and adapting to diverse student needs. However, the effectiveness of clustering depends on its interpretability, which becomes challenging with high-dimensional data. Existing clustering approaches often neglect individual differences in feature importance and rely on a homogenized feature set. Addressing this gap, we introduce \texttt{Interpret3C} (Interpretable Conditional Computation Clustering), a novel clustering pipeline that incorporates interpretable neural networks (NNs) in an unsupervised learning context. This method leverages adaptive gating in NNs to select features for each student. Then, clustering is performed using the most relevant features per student, enhancing clusters' relevance and interpretability. We use \texttt{Interpret3C} to analyze the behavioral clusters considering individual feature importances in a MOOC with over $5,000$ students. This research contributes to the field by offering a scalable, robust clustering methodology and an educational case study that respects individual student differences and improves interpretability for high-dimensional data.

\end{abstract}

\section{Introduction}
Clustering a student population into educationally relevant groups facilitates a deeper understanding of student behaviors and learning patterns, enabling educators to optimize curriculum design and implement group interventions \cite{effenberger2021interpretable,mejia-domenzain_evolutionary_2022,mejia-domenzain_identifying_2022,peng2022}. The effectiveness of these strategies hinges on the interpretability of the clusters as it directly influences the quality of insights educators can extract from them \cite{effenberger2021interpretable,swamy2023future}. In large-scale educational settings such as Massive Open Online Courses (MOOCs), where individualized attention is challenging due to the sheer number of participants, clustering provides a feasible approach to understand and address the needs of different student groups. 

In these online contexts, abundant and varied features have been used to study student behavior \cite{DBLP:conf/edm/AkpinarRA20,DBLP:conf/ectel/BoroujeniSKLD16,chen2020utilizing,DBLP:conf/aied/LalleC20,DBLP:journals/eait/LemayD20,marrascan,DBLP:conf/aied/MbouzaoDS20,DBLP:journals/eait/MubarakCA21,DBLP:journals/tlt/WanLYG19}. A broad spectrum of features is necessary for a comprehensive study of complex student behavior. However, it simultaneously poses substantial challenges in terms of robustness and interpretability due to the \textit{curse of dimensionality}. In high-dimensional spaces, the sparsity of data makes most distance measures less effective, leading to less robust results \cite{hancer2020survey}. In addition, a large number of features can lead to intricate yet less intuitive clustering outputs, thereby complicating the translation of these results into practical educational applications, such as supporting educators to identify and address areas where groups of students may require assistance.

To target the issues of interpretability and robustness, previous works have used a subset of expert-selected features \cite{choi23,mejia-domenzain_identifying_2022,mejia-domenzain_evolutionary_2022}, and data-driven feature selection methods \cite{effenberger2021interpretable,peffer2020trace} before clustering.
The learning science-driven approach has been to select a limited number of setting-specific features  \cite{DBLP:conf/ectel/BoroujeniSKLD16,choi23,marrascan,mejia-domenzain_evolutionary_2022,mejia-domenzain_identifying_2022}. For example, \cite{choi23} followed the domain modeling step of the evidence-centered design framework to engineer the features used for clustering. However, the process of hand-picking features is heavily reliant on the expertise and perspectives of the researchers or educators involved.  This reliance can inadvertently introduce subjective biases, as the selected features may reflect the specific hypotheses or expectations of the individuals involved, rather than the full spectrum of student behaviors and learning patterns. On the other hand, there are multiple data-driven approaches for feature selection in unsupervised settings, including filter, wrapper, hybrid, and embedded approaches \cite{hancer2020survey}. A significant limitation within these approaches is that the features are chosen averaging over all students, thereby neglecting individual differences in feature importance. This oversight results in an aggregation process that identifies features beneficial on a global scale but fails to account for unique student characteristics.

Addressing this challenge, neural networks (NNs) could be adapted to act as \textit{embedded} methods to perform individual feature selection as part of their training process. Compared to traditional ML methods, NNs are exceptionally good at capturing complex, non-linear relationships and interactions between features. This characteristic could enable NNs to identify important features in datasets where the significance of a feature is not obvious or is dependent on interactions with other features. Moreover, NNs are well suited for high-dimensional data, as they are robust to irrelevant or redundant features. However, NNs face interpretability issues due to their complex, multi-layered ``black box'' operations \cite{peng2022,swamy2024interpretcc,swamy2023future}. To address this issue, \cite{swamy2024interpretcc} proposed an interpretable NN architecture (InterpretCC) that uses a dynamic feature mask to enforce sparsity regularization on the number of input features. This method enables feature importances vectors per student without compromising classification or regression accuracy. 

While interpretable-by-design NNs offer a promising solution for dealing with high-dimensional data and individualized feature selection, their application has predominantly been in supervised tasks \cite{DBLP:journals/eait/MubarakCA21,swamy2024interpretcc,swamy2023future}. The extension of these methods to unsupervised settings, particularly for interpretable clustering, remains unexplored. This could address the existing gap in the consideration of individual differences in clustering. Traditional feature selection techniques, typically employed before clustering, opt for a global feature selection strategy \cite{hancer2020survey}. Such a strategy creates a homogenized feature set that fails to recognize and incorporate the distinctive attributes of individual students. 

This work bridges the aforementioned gaps by developing \texttt{Interpret3C} (Interpretable Conditional Computation Clustering), a clustering pipeline that leverages the strengths of interpretable NNs in an unsupervised learning context. This approach is tailored to adaptively select features per student, thereby facilitating easier and more meaningful interpretation of clusters without compromising the quality and robustness of the clustering process. This is achieved through a deep feature selection method where the individual feature importance masks are extracted from interpretable NNs trained to predict academic performance. Clustering is then performed on the important features. We evaluate our pipeline on a large MOOC with over 5,000 enrolled students and hundreds of thousands of interactions to address the following research question: What kind of student behavioral clusters are identified through inherently interpretable clustering?

\section{Methodology}
\label{sec:method}
We contribute an inherently interpretable clustering pipeline\footnote{\url{https://github.com/epfl-ml4ed/interpretable-clustering/}}. Our approach involves the adaptive selection of the most pertinent features for individual students, and uses only these identified features for the clustering process. 

\vspace{-3mm}
\subsection{Student Interactions and Feature Extraction}
We begin by collecting clickstream data from MOOCs on student interactions from videos and problems, and their associated actions, which include video-related actions. We extract 45 behavioral features from the student interactions\footnote{The full list of features is available at \url{https://github.com/epfl-ml4ed/interpretable-clustering/blob/main/docs/features-description.pdf}} that have been found to be predictive in MOOCs \cite{DBLP:conf/ectel/BoroujeniSKLD16,chen2020utilizing,DBLP:conf/aied/LalleC20,marrascan,swamy2022meta}. The feature sets encompass a wide array of online learning behaviors, including clickstream patterns, study regularity, quiz performance, and video interaction metrics. \cite{DBLP:conf/ectel/BoroujeniSKLD16} evaluates study pattern regularity and \cite{chen2020utilizing} explores clickstream activities. Video interactions in MOOCs are analyzed in detail by \cite{DBLP:conf/aied/LalleC20}, while \cite{marrascan} focuses on features related to quiz performance. 

Each feature is computed on a weekly basis, resulting in a time series vector for every student. This transforms each student's behavior into a multivariate time series, where one dimension represents the progression of weeks and the other encapsulates the diverse set of features. We then use unit-norm scaling to normalize the features, preventing vanishing or exploding gradients in the deep feature selection step. 

\begin{figure}[t]
    \centering
    \includegraphics[clip, width=1\textwidth]{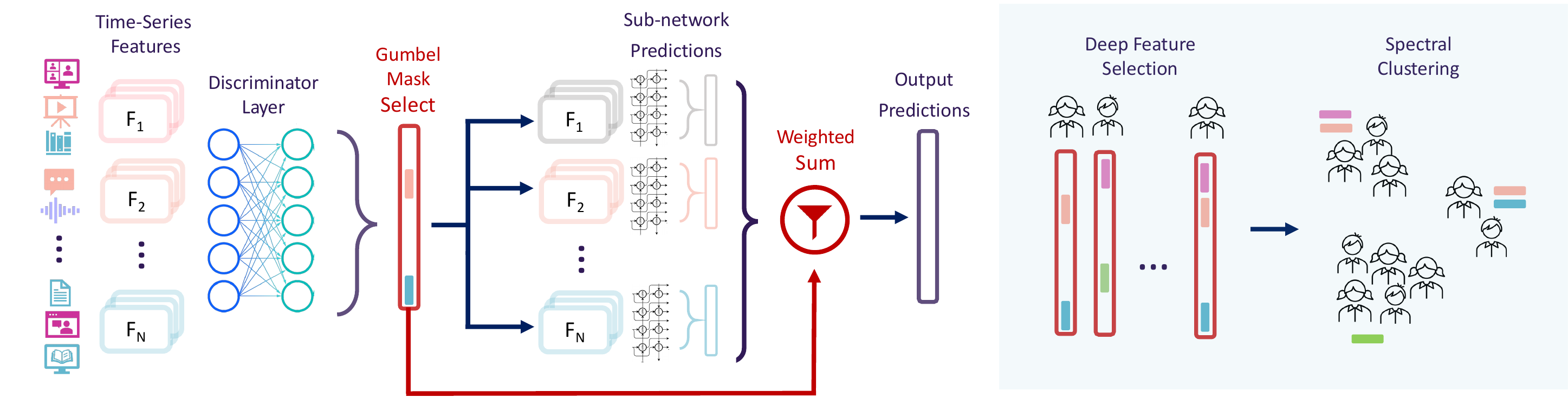}
    \caption{\texttt{Interpret3C} pipeline with deep feature selection and clustering.}
    \label{fig:fg_arch}
    \vspace{-4mm}
\end{figure}

\vspace{-3mm}
\subsection{Deep Feature Selection}
We extract the most important features for each student using an interpretable NN architecture. We define feature importance as the degree to which individual features contribute to the predictive accuracy of the model. Thus, to extract the individual feature importance masks, we train NNs on the supervised task of predicting students' academic performance (pass or fail).

The architecture, an extension of the predictive model presented by \cite{swamy2024interpretcc}, leverages a feature gating mechanism that dynamically selects relevant input features for making predictions. As shown in Fig.~\ref{fig:fg_arch}, the features initially pass through discriminator layers, followed by a sigmoid function that produces a feature mask. Features with a sigmoid output of 0.5 or higher are considered activated. Each activated feature is then processed by a dedicated BiLSTM sub-network to predict the student's likelihood of passing or failing. The overall prediction is the average of these individual predictions, weighted by the feature activations.

To encourage interpretability through sparsity, we integrate an annealed mean-squared error regularization on the feature mask. Moreover, the incompatibility of discrete feature gating with backpropagation is overcome by using the Gumbel-SoftMax technique. This technique approximates discrete choices through differentiable functions by adding Gumbel noise and applying softmax as done in \cite{swamy2024interpretcc}.

\vspace{-3mm}
\subsection{Clustering}
To obtain the clusters, we use as input the masked feature matrix per student containing only the important feature vectors. For instance, if a student $s$ has a feature mask $m_{s} = (0,1,1)$ and their corresponding feature matrix is $F_{s} = (f_{s}^{1}, f_{s}^{2}, f_{s}^{3})$, with $f_{s}^{i}$ representing the time series vector for each feature $i$, the resulting masked feature matrix would be $F_{m} = (-, f_{s}^{2}, f_{s}^{3})$. Next, for each feature $f$, we calculate $D_{f}$ the Euclidean pairwise distance between students' time series. To calculate distances between vectors in the presence of incomplete data, we impute a vector of zeros. This allows non-missing features to have a stronger contribution to the distance calculation. Following, we compute $D$ as the average of all the features' distance matrix ($D_{f}$). Next, we apply a Gaussian Kernel to compute $S_{f}$, the similarity matrix of $D_{f}$. We use Spectral Clustering to cluster $S_{f}$ with the number of clusters $n$ as a hyper-parameter. We considered from 3 to 10 clusters and choose $n$ according to the Eigengap heuristic as suggested in \cite{von2007tutorial}. We deliberately choose $n$ to be greater than two to prevent replicating the outcomes of the binary classification task (pass or fail) and obtain additional insights into student behavior. 

\section{Experimental Evaluation} 
In our analysis, we trained an \texttt{Intepret3C} pipeline to examine the clusters obtained over thousands of students in a MOOC on Digital Signal Processing. In particular, the course is a Computer Science Master's course taught in English with $5{,}611$ active students for 10 weeks. The choice of this course was based on its prior use in other studies \cite{marrascan,swamy2022meta,swamy2022evaluating}.

\begin{figure}[t]
    \centering
    \includegraphics[width=\textwidth]{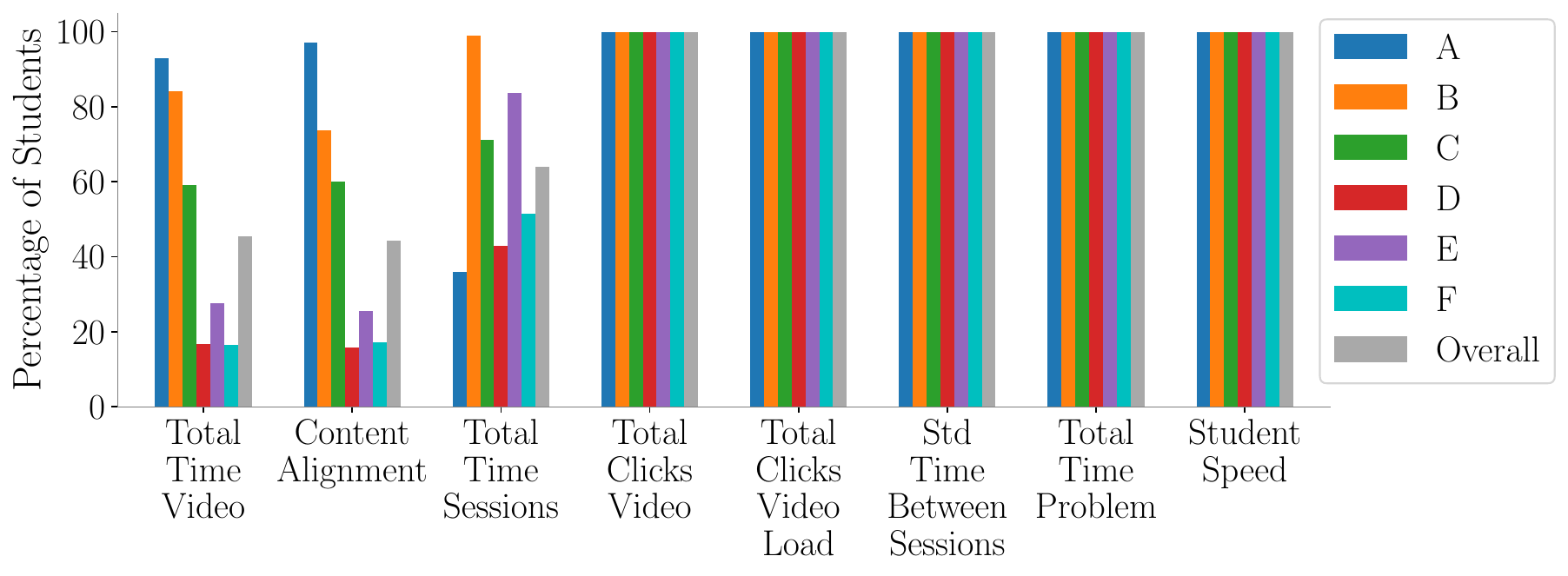}
    \vspace{-5mm}
    \caption{For each of 8 important features (\textit{x-axis}), the percentage of students (\textit{y-axis}) from each cluster (\textit{color}) that selected the feature as important.}\label{fig:percentage_students_per_cluster}
\end{figure}

\begin{figure}[t]
    \centering
    \includegraphics[width=\textwidth]{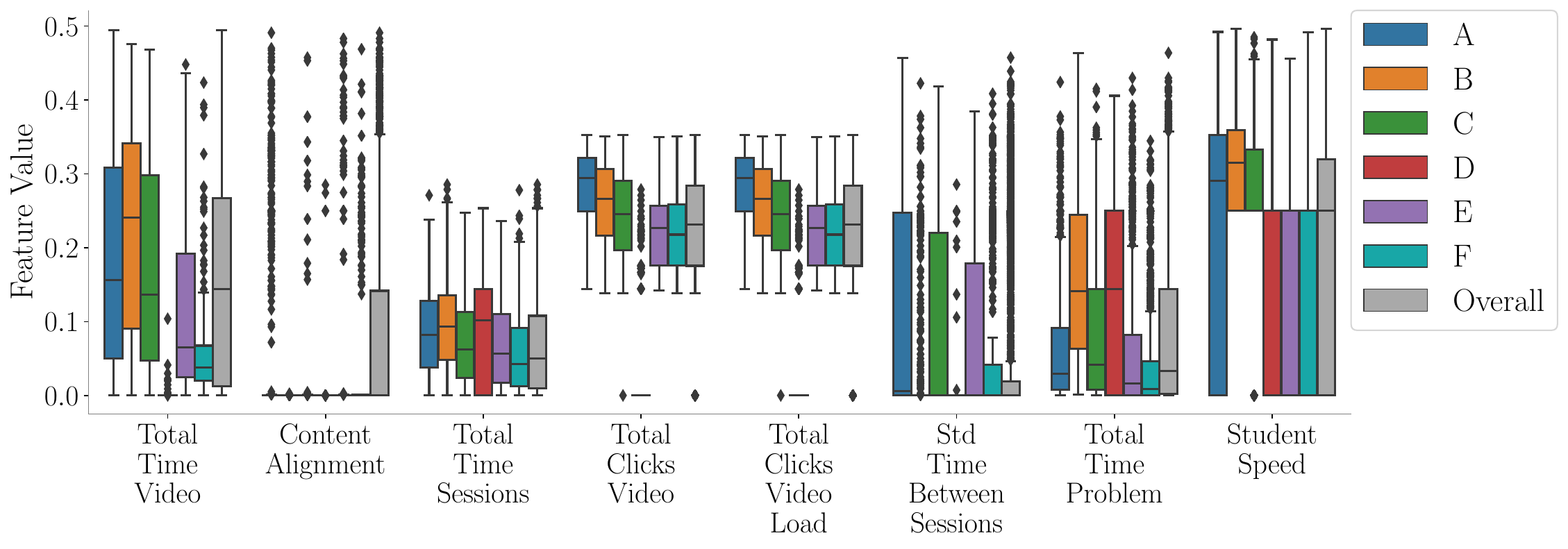}
        \vspace{-5mm}
    \caption{For each of 8 important features (\textit{x-axis}), the feature value distribution (\textit{y-axis}) from each cluster (\textit{color}).}
    \vspace{-2mm}\label{fig:feature_value_per_cluster}
\end{figure}

We use the first four weeks in an early prediction setting to identify six behavioral clusters labeled A through F and ordered according to academic outcomes. Cluster A is the group with the highest percentage of students passing the course, while students in F have the highest failure rate. The distribution of students across these clusters is as follows: Cluster A contains 786 students, equating to 14\% of the course's population, with a passing rate of 60\%. Clusters B through F comprise 15\%, 14\%, 14\%, 18\%, 25\% of the student body, respectively, with passing rates of 42\%, 31\%, 17\%, 15\%, 11\%.

Fig.~\ref{fig:percentage_students_per_cluster} illustrates the distribution of important features within each cluster. For each cluster, it shows the percentage of students for whom each feature was deemed relevant. One initial observation is that out of the $45$ original features, only eight were selected as important for at least one student. Out of these eight features, five emerged as significant across all clusters. The features are related to video activities (\textit{Total Clicks Video Load}, \textit{Total Clicks Video}), quiz interactions (\textit{Time in Problem Sum}, and \textit{Student Speed}) and regularity patterns (\textit{Time Between Sessions Std}). The other three features (\textit{Total Time Video}, \textit{Content Alignment} and \textit{Total Time Session}) vary considerably between clusters. Only in the best-performing clusters (A, B, and C), \textit{Total Time Video} and \textit{Content Alignment} were selected as important for the majority of the students.

To gain a deeper understanding of the differences between the clusters, Fig.~\ref{fig:feature_value_per_cluster} presents a comparison of the average feature values for each important feature across clusters. It also includes the ``Overall" category, which shows the average feature values for all students. Different from Fig.~\ref{fig:percentage_students_per_cluster} where there were slight variations in the percentages of feature importances, in Fig.~\ref{fig:feature_value_per_cluster} the distributions vary considerably between clusters. This variation suggests that while a feature may be considered important across clusters, the degree to which it is manifested in student behavior could differ. One specific feature of interest is ``Content Alignment'', which measures schedule adherence. In Fig.~\ref{fig:feature_value_per_cluster}, this feature consistently exhibits one of the highest levels of variability, with both the 25th and 75th percentiles at zero for all clusters. This indicates that the vast majority of students did not complete the video content within the designated timeframe. This contrasts with the \textit{Overall} category distribution of the feature where the definition of outlier is above 0.35 (instead of zero). Thus, it seems that \textit{Content Alignment} is mostly considered as an important feature for the students who are behind schedule. For example, in Clusters A and B, it was an important feature for more than 70\% of the students. Interestingly, in Cluster B, all students had a value of zero for this feature, while for Cluster A, there were multiple outliers with values greater than zero. This suggests that while this feature was important for both clusters, students in Cluster B were universally off-track, whereas in Cluster A, a subset of the students managed to adhere to the schedule.

An additional observation is the difference in engagement between the best-performing clusters (A and B). As seen in Fig.~\ref{fig:feature_value_per_cluster}, students in Cluster B dedicate more time to videos and problem-solving than those in Cluster A. However, students in Cluster A exhibit more active engagement, as evidenced by lower \textit{Student Speed} values related to quiz attempt frequency as well as higher values in video-related features such as \textit{Total Video Clicks}. 

Furthermore, Fig.~\ref{fig:feature_value_per_cluster} reveals a general downward trend in feature values, with high-performing clusters (A, B, and C) having higher values compared to low-performing clusters (E and F). Cluster D, however, deviates from this trend by exhibiting the lowest median values (zero) for all video-related features. Despite this, students in Cluster D are not disengaged from the course as they exhibit the longest time online, attributed primarily to their extensive interaction with quizzes (\textit{Total Time Problems}).
The passing rate for Cluster D is 16.8\%, which is marginally higher than the 15.2\% passing rate of Cluster E. Nevertheless, their interaction with the course material was very different. The students in Cluster D spent a lot of time online interacting with the quizzes, while the students in Cluster E showed lower levels of engagement across both videos and quizzes. 

\section{Discussion and Conclusion}

In this work, we presented a novel methodology and initial evaluation for an interpretable-by-design clustering pipeline, \texttt{Interpret3C}.

We performed an in-depth cluster analysis based on student behavior from the first four weeks of a large MOOC course. Our pipeline revealed six diverse behavioral clusters, each characterized by varying levels of interaction with course materials and platform features. The contrast between Cluster D, highly engaged in quizzes, and Cluster E, generally disengaged, illustrates that while the outcomes may appear similar, the underlying behaviors can differ significantly; thus, effective interventions should address the specific needs of each group. Moreover, we found five features important for all students and three features with a varying percentage of importance across clusters. Global feature selection methods \cite{effenberger2021interpretable,hancer2020survey,peffer2020trace} could have also identified the five universal relevant features but would likely have overlooked the nuanced variations in the importance of \textit{Total Time Video}, \textit{Content Alignment}, and \textit{Total Time Sessions} across different clusters. 

The quality and relevance of the selected features depends strongly on 1) the signal of the input features and 2) the performance of the discriminator network and subsequent time series networks, which require a relatively complex architecture. Although the utilization of these resources for feature selection may initially appear excessive, the number of parameters is much smaller than a LLM or other models used for educational deep learning \cite{DBLP:journals/eait/MubarakCA21,swamy2022meta}. In these settings, the training can serve a double purpose: to predict and also to learn about the group behaviors in the class. Furthermore, the deep feature selection process has a bias towards the initial predictive task of the discriminator NN. In our case, the selection mechanism identifies important features related to academic outcomes. This bias towards the initial task is not inherently negative, as the prediction task can be adapted to fit other educational goals (e.g., retention or self-regulated learning skills). Including the outcome measure in unsupervised settings mirrors that of other methodologies, such as supervised PCA and sparse supervised CCA \cite{witten2009}, helping to identify features associated with the outcomes and correlations between datasets. Extending the Interpret3C approach to other tasks and interpreting the clusters over a more generalizable set of courses is left as future work.

In conclusion, \texttt{Interpret3C} introduces an innovative clustering pipeline by emphasizing the interpretability of high-dimensional data and respecting individual student differences. Through adaptive feature selection, our approach enhances the relevance and clarity of clustering outcomes. Our findings illustrate the potential of \texttt{Interpret3C} to identify insightful clusters in MOOC environments and paves the way for more effective and personalized group interventions.

\vspace{2mm} \noindent \textbf{Acknowledgements}. We kindly thank  the Swiss State Secretariat for Education, Research and
Innovation (SERI) for supporting this project.

\bibliographystyle{splncs04}
\bibliography{references}

\end{document}